\documentstyle[aps,amssymb]{revtex}

\begin{document}
\author{Yong Chen, Ying Hai Wang and Kong Qing Yang}
\title{The macroscopic dynamics in separable neural networks\thanks{%
Published in Phys. Rev.E 63, 041901(2001)}}
\address{Department of Physics, Lanzhou University, Lanzhou, Gansu, 730000, China}
\maketitle

\begin{abstract}
The parallel dynamics is given in the case of neural networks with separable
coupling through starting from Coolen-Sherrington (CS) theory. It is shown
that this retrieve dynamics as is the case of sequential evolution in the
postulate of away from saturation and finite temperature. The finite-size
effects is governed by a homogeneous Markov process, which differs from the
time-dependent Ornstein-Uhlenbeck process in sequential dynamics.\newline
PACS number(s): 87.10.+e, 75.10.Nr, 02.50.+s
\end{abstract}

\bigskip 

Ising spin models for neural networks have made a more and more significant
contribution to understand the information processing in biotic nervous
system since the pioneering work by Little\cite{little 1974}, Hopfield\cite
{hopfield 1982} and Amit {\it et al}\cite{amit 1985}. Moreover, both far
from and near saturation, starting from the theory of equilibrium
statistical mechanics of spin-glass-like systems, a survey of the properties
of the Hopfiled model is given by Amit, Gutfreund and Sompolinsky\cite{amit
1985}\cite{amit 1987}, in the case of symmetric connections. It means that a
static analysis is enough to investigate the networks with symmetric
coupling. However, when one deals with networks with asymmetric connections,
which are ubiquitous in real neurons of living systems\cite{eccles 1977}, it
becomes more important to consider the dynamics rather than the equilibrium
properties of networks. Recently, by Markovian sequential dynamics, Coolen 
{\it et al.} developed a series of analytic scheme for networks with
separable coupling which include symmetric and asymmetric \cite{coolen 1988}%
\cite{coolen 1992-1}\cite{coolen 1993}. It is clear that there exist two
types of deterministic dynamics, sequential and parallel, in evolution of
systems. One may question: (1) How about the macroscopic description of this
systems with synchronous case? There are difference between both case? (2)
Additionally, how are the finite-size effect in parallel case comparing with
it in sequential case presented by Castellanos, Coolen and Viana\cite
{castellanos 1998}?

Since studied systems cover up the case of asymmetric conjunctions, we
resort to stochastic analysis. We take up the Markovian dynamics, which is
considered to have advantage that it not only provides a simple description
of stochastic process but also enable to deal with nonequilibrium properties%
\cite{Haken 1975}. So, in this work we offer the foregoing issues based upon
the way of Markov process analysis and CS theory.

As usual, in Ising spin model of a neural networks constructed by $N$
neurons, the states of the $i$-th neuron in time $t$ is described by $%
s_i\left( t\right) \in \left\{ 1,-1\right\} $. Moreover, the networks has
stored $p$ sets of patterns $\xi _i^\mu \in \left\{ 1,-1\right\} \left( \mu
=1,2,\ldots ,p,{\rm \quad }i=1,2,\ldots ,N\right) $ which are embedded for
the purpose of associative memory retrieval through the synaptic connections
with the Hebb learning rule taken into account\cite{hebb 1957}. We take the
separable interaction matrices ${\bf J}$ to be\cite{coolen 1988}\cite{shiino
1990} 
\begin{equation}
J_{ij}=\frac 1N\sum\limits_{\mu ,\nu =1}^P\xi _i^\mu A_{\mu \nu }\xi _j^\nu
\label{1}
\end{equation}
where the ${\bf A}$ is $p\times p$ matrix representing all kinds of
separable conjunctions ${\bf J}$ in common sense. In there, the coupling
matrix, both symmetric and asymmetric, can be represented by the form of
definition (\ref{1}) in most case. Further more, the synchronous evolution
dynamics of these systems can be defined as: 
\begin{equation}
s_i\left( t+\triangle t\right) ={\normalsize sgn}\left(
\sum\limits_{j=1}^NJ_{ij}s_j\left( t\right) \right)  \label{2}
\end{equation}
where $\triangle t$ is the length of each time step. In there, the threshold
of every nerve cell is set as constant one, and the update function is step
function ${\rm sgn}(x)$ which its value is $1$ for $x\geq 0$ and $-1$ for $%
x<0$.

Now, according to a continuous Markov process\cite{nelken 1988}, the
probability $P_{t+\vartriangle t}\left( {\bf s}\right) $ of finding the
system in state ${\bf s=}\left( s_1,s_2,\ldots ,s_N\right) ^T$ at time $%
t+\triangle t$ is given by: 
\begin{equation}
P_{t+\vartriangle t}\left( {\bf s}\right) =\sum\limits_{{\bf s}^{\prime
}}W\left( {\bf s}^{\prime }\rightarrow {\bf s}\right) P_t\left( {\bf s}%
^{\prime }\right)  \label{3}
\end{equation}
where $W\left( {\bf s}^{\prime }\rightarrow {\bf s}\right) $ is the
so-called transfer probability or jump probability, which means finding the
system in state ${\bf s}$ at time $t+\triangle t$ given it is in state ${\bf %
s}^{\prime }$ at $t$ time. Because of the normalization condition 
\[
\sum\limits_{{\bf s}}W\left( {\bf s}^{\prime }\rightarrow {\bf s}\right)
=\sum\limits_{{\bf s\neq s}^{\prime }}W\left( {\bf s}^{\prime }\rightarrow 
{\bf s}\right) +W\left( {\bf s}^{\prime }\rightarrow {\bf s}^{\prime
}\right) =1 
\]
it is easy to get the following discrete mapping 
\begin{equation}
P_{t+\triangle t}\left( {\bf s}\right) =P_t\left( {\bf s}\right)
+\sum\limits_{{\bf s}^{\prime }\neq {\bf s}}W\left( {\bf s}^{\prime
}\rightarrow {\bf s}\right) P_t\left( {\bf s}^{\prime }\right) -\sum\limits_{%
{\bf s}^{\prime }\neq {\bf s}}W\left( {\bf s}\rightarrow {\bf s}^{\prime
}\right) P_t\left( {\bf s}\right)  \label{4}
\end{equation}
When the time step $\triangle t$ go to zero, one obtains a continuous time
version of digital master equations (\ref{4}): 
\begin{equation}
\frac{dP_t\left( {\bf s}\right) }{dt}=\sum\limits_{{\bf s}^{\prime }\neq 
{\bf s}}w\left( {\bf s}^{\prime }\rightarrow {\bf s}\right) P_t\left( {\bf s}%
^{\prime }\right) -\sum\limits_{{\bf s}^{\prime }\neq {\bf s}}w\left( {\bf s}%
\rightarrow {\bf s}^{\prime }\right) P_t\left( {\bf s}\right)  \label{5}
\end{equation}
with $w\left( {\bf s}^{\prime }\rightarrow {\bf s}\right)
=\lim\limits_{\triangle t\rightarrow 0}\frac{W\left( {\bf s}^{\prime
}\rightarrow {\bf s}\right) }{\triangle t}$. In there, $w\left( {\bf s}%
^{\prime }\rightarrow {\bf s}\right) $ represents the density of unit time
for transfer probability (it is equal to transition rate), which the states
of system change from ${\bf s}^{\prime }$ to ${\bf s}$, in the interval
between $t$ and $t+\triangle t$. For parallel evolutionary process,
reflecting upon the conclusion in biological statistics, the transition rate
can be given by\cite{shiino 1990}\cite{peretto 1986} 
\begin{equation}
w\left( {\bf s}^{\prime }\rightarrow {\bf s}\right) =\prod\limits_i^N\frac 12%
\left( 1+s_i\tanh \left( \beta h_i\left( {\bf s}^{\prime }\right) \right)
\right)  \label{6}
\end{equation}
where $h_i\left( {\bf s}^{\prime }\right) \equiv \sum\nolimits_jJ_{ij}s_j$
is local fields of a stochastic alignment of the spins, and $\beta \equiv 
\frac 1T$ denotes a measure of the inverse magnitude of the amount of noise
affecting the neurons, acting as the role of temperature in analogy to
thermodynamic spin systems.

Generally, we are interested in the problem of macroscopic features rather
more than the microscopic details of networks. Sequentially, one introduce
any set of linear order variables $\Omega _k$\cite{coolen 1988}\cite{parisi
1980}: 
\begin{equation}
\Omega _\mu \left( {\bf s}\right) =\frac 1N\sum\limits_j\xi _j^\mu s_j\qquad
P_t\left( {\bf \Omega }\right) =\sum\limits_{{\bf s}}P_t\left( {\bf s}%
\right) \delta \left( {\bf \Omega }-{\bf \Omega }\left( {\bf s}\right)
\right) \qquad \mu =1,2,\ldots ,p  \label{7}
\end{equation}
where $P_t\left( {\bf \Omega }\right) $ denotes the probability of finding $%
{\bf \Omega }$ in time $t$. Following up the above definition and the Eq. (%
\ref{5}), it is easy to get 
\[
\frac{dP_t\left( {\bf \Omega }\right) }{dt}=\sum\limits_{{\bf s}%
}\sum\limits_{{\bf s}^{\prime }\neq {\bf s}}w\left( {\bf s}^{\prime
}\rightarrow {\bf s}\right) P_t\left( {\bf s}^{\prime }\right) \delta \left( 
{\bf \Omega }-{\bf \Omega }\left( {\bf s}\right) \right) -\sum\limits_{{\bf s%
}}\sum\limits_{{\bf s}^{\prime }\neq {\bf s}}w\left( {\bf s}\rightarrow {\bf %
s}^{\prime }\right) P_t\left( {\bf s}\right) \delta \left( {\bf \Omega }-%
{\bf \Omega }\left( {\bf s}\right) \right) 
\]
Clearly, the above master equation can be replaced by 
\begin{equation}
\frac{dP_t\left( {\bf \Omega }\right) }{dt}=\sum\limits_{{\bf s}%
}\sum\limits_{{\bf s}^{\prime }}w\left( {\bf s}\rightarrow {\bf s}^{\prime
}\right) P_t\left( {\bf s}\right) \left( \delta \left( {\bf \Omega }-{\bf %
\Omega }\left( {\bf s}^{\prime }\right) \right) -\delta \left( {\bf \Omega }-%
{\bf \Omega }\left( {\bf s}\right) \right) \right)  \label{8}
\end{equation}
Following after CS theory\cite{coolen 1992-1}, we introduce any function $%
\Phi \left( {\bf \Omega }\right) $ and its average is $\left\langle \Phi
\left( {\bf \Omega }\right) \right\rangle _t\equiv \int d{\bf \Omega }%
P_t\left( {\bf \Omega }\right) \Phi \left( {\bf \Omega }\right) $. Its time
differential is 
\begin{eqnarray}
\left\langle \Phi \left( {\bf \Omega }\right) \right\rangle _t
&=&\sum\limits_{{\bf s}}\sum\limits_{{\bf s}^{\prime }}w\left( {\bf s}%
\rightarrow {\bf s}^{\prime }\right) P_t\left( {\bf s}\right) \left[ \Phi
\left( {\bf \Omega }\left( {\bf s}^{\prime }\right) \right) -\Phi \left( 
{\bf \Omega }\left( {\bf s}\right) \right) \right]  \nonumber \\
&=&\sum\limits_{{\bf s}}\sum\limits_{{\bf s}^{\prime }}w\left( {\bf s}%
\rightarrow {\bf s}^{\prime }\right) P_t\left( {\bf s}\right)
\sum\limits_{n=1}\frac 1{n!}\sum\limits_{\mu _1=1}^n\sum\limits_{\mu
_2=1}^n\cdots \sum\limits_{\mu _p=1}^n  \nonumber \\
&&\left[ {\bf \Omega }\left( {\bf s}^{\prime }\right) -{\bf \Omega }\left( 
{\bf s}\right) \right] ^{\mu _1+\mu _2+\cdots +\mu _p}\frac{\partial ^n\Phi
\left[ {\bf \Omega }\left( {\bf s}\right) \right] }{\partial \Omega _{\mu
_1}\partial \Omega _{\mu _2}\cdots \partial \Omega _{\mu _p}}  \label{9}
\end{eqnarray}
where there exist appendant condition that the sum $\sum_i^p\mu _i=n$.
Inserting the unit operator $\int d{\bf \Omega }\delta \left( {\bf \Omega }-%
{\bf \Omega }\left( {\bf s}\right) \right) $ and then performing partial
integrations yields 
\begin{eqnarray}
\frac{dP_t\left( {\bf \Omega }\right) }{dt} &=&-\sum\limits_{{\bf s}%
}\sum\limits_{{\bf s}^{\prime }}\sum\limits_{n=1}\frac 1{n!}\sum\limits_{\mu
_1=1}^n\sum\limits_{\mu _2=1}^n\cdots \sum\limits_{\mu _p=1}^n\left[ {\bf %
\Omega }\left( {\bf s}^{\prime }\right) -{\bf \Omega }\left( {\bf s}\right)
\right] ^{\mu _1+\mu _2+\cdots +\mu _p}  \nonumber \\
&&\frac{\partial ^nw\left( {\bf s}\rightarrow {\bf s}^{\prime }\right)
P_t\left( {\bf s}\right) \delta \left( {\bf \Omega }-{\bf \Omega }\left( 
{\bf s}\right) \right) }{\partial \Omega _{\mu _1}\partial \Omega _{\mu
_2}\cdots \partial \Omega _{\mu _p}}  \label{10}
\end{eqnarray}
In a word, this equation is the forms with Kramers-Moyal-like expansion for
the master Eq. (\ref{8}) for the probability of the introduced order
variables (pattern overlaps). Under this equation with expansion forms, one
can study the properties of pattern overlaps, describing the macroscopic
behavior, with its lowest order term.

In fact, the Eq. (\ref{10}) is equal to 
\begin{eqnarray}
\frac{dP_t\left( {\bf \Omega }\right) }{dt} &=&-\sum\limits_\mu \frac %
\partial {\partial \Omega _\mu }\left\{ \sum\limits_{{\bf s}}\sum\limits_{%
{\bf s}^{\prime }}\left[ {\bf \Omega }\left( {\bf s}^{\prime }\right) -{\bf %
\Omega }\left( {\bf s}\right) \right] w\left( {\bf s}\rightarrow {\bf s}%
^{\prime }\right) P_t\left( {\bf s}\right) \delta \left( {\bf \Omega }-{\bf %
\Omega }\left( {\bf s}\right) \right) \right\}  \nonumber \\
&&+\sum_{m\geq 2}{\cal O}\left( NP\left( \frac 2N\right) ^m\right)
\label{11}
\end{eqnarray}
Consequently, at the limit case of finite temperature and far away
saturation, this becomes 
\begin{equation}
\frac{dP_t\left( {\bf \Omega }\right) }{dt}=-\sum\limits_k\frac \partial {%
\partial \Omega _k}\left\{ \sum\limits_{{\bf s}}\left[ \sum\limits_{{\bf s}%
^{\prime }}\left[ {\bf \Omega }\left( {\bf s}^{\prime }\right) -{\bf \Omega }%
\left( {\bf s}\right) \right] w\left( {\bf s}\rightarrow {\bf s}^{\prime
}\right) \right] P_t\left( {\bf s}\right) \delta \left( {\bf \Omega }-{\bf %
\Omega }\left( {\bf s}\right) \right) \right\}  \label{12}
\end{equation}
we denote $\vartriangle _i=\frac 12\left( 1-s_i\tanh \left( \beta h_i\left( 
{\bf s}\right) \right) \right) $ and $\vartriangle _i^{\prime }=\frac 12%
\left( 1+s_i\tanh \left( \beta h_i\left( {\bf s}\right) \right) \right) ,$
corresponding with $w\left( s_i\rightarrow -s_i\right) $ and $w\left(
s_i\rightarrow s_i\right) $ respectively. According to the definitions of
order parameters ${\bf \Omega }$ and the coupling matrix ${\bf J}$, the
stochastic local field $h_i\left( {\bf s}\right) $ is equal to 
\[
h_i\left( {\bf s}\right) =\frac 1N\sum\limits_jJ_{ij}s_j=\sum_{\mu ,\nu }\xi
_i^\mu A_{\mu \nu }\left( \frac 1N\sum_j\xi _j^\nu s_j\right) ={\bf \xi }%
_i\cdot {\bf A\Omega } 
\]
Then, the transition rate can be rewritten as 
\[
\vartriangle _i=\frac 12\left( 1-s_i\tanh \left( \beta {\bf \xi }_i\cdot 
{\bf A\Omega }\right) \right) \qquad \vartriangle _i^{\prime }=\frac 12%
\left( 1+s_i\tanh \left( \beta {\bf \xi }_i\cdot {\bf A\Omega }\right)
\right) 
\]
Therefore, for the terms of $\left[ \cdots \right] $ in the right side of
Eq. (\ref{12}), from definition (\ref{6}) and (\ref{7}), we get 
\begin{eqnarray}
\sum\limits_{{\bf s}^{\prime }}\left[ {\bf \Omega }\left( {\bf s}^{\prime
}\right) -{\bf \Omega }\left( {\bf s}\right) \right] w\left( {\bf s}%
\rightarrow {\bf s}^{\prime }\right) &=&-\frac 2N\left( \sum\limits_i^N\frac{%
{\bf \xi }_is_i\frac{\vartriangle _i}{\vartriangle _i^{\prime }}}{1+\frac{%
\vartriangle _i}{\vartriangle _i^{\prime }}}\right) \left(
\prod\limits_j^N\left( 1+\frac{\vartriangle _j}{\vartriangle _j^{\prime }}%
\right) \right) \left( \prod\limits_k^N\vartriangle _k^{\prime }\right) 
\nonumber \\
&=&-2\frac 1N\sum\limits_i^N{\bf \xi }_is_i\vartriangle _i  \nonumber \\
&=&-{\bf \Omega }\left( {\bf s}\right) +\frac 1N\sum\limits_i^N{\bf \xi }%
_i\tanh \left( \beta {\bf \xi }_i\cdot {\bf A\Omega }\right)
\end{eqnarray}
Substituting the above relation into Eq. (\ref{12}) yields 
\begin{equation}
\frac d{dt}P_t\left( {\bf \Omega }\right) =-\sum\limits_\mu \frac \partial {%
\partial \Omega _\mu }\left\{ P_t\left( {\bf \Omega }\right) \left( -{\bf %
\Omega }\left( {\bf s}\right) +\frac 1N\sum\limits_i^N{\bf \xi }_i\tanh
\left( \beta {\bf \xi }_i\cdot {\bf A\Omega }\right) \right) \right\}
\label{13}
\end{equation}
On the basis of Markov process theory\cite{Gardiner 1983}, obviously, the
deterministic Liouville forms of Eq. (\ref{13}) is 
\begin{equation}
\frac d{dt}\Omega _\mu =\lim\limits_{N\rightarrow \infty }\left( \frac 1N%
\sum\limits_i\xi _i^\mu \tanh (\beta {\bf \xi }_i\cdot {\bf A\Omega }%
)-\Omega _\mu \right)  \label{14}
\end{equation}
with the initial value can be set as ${\bf \Omega }\left( 0\right) ={\bf %
\Omega }_0=\frac 1N\sum_i{\bf \xi }_is_i\left( 0\right) $.

Comparing the Eq. (\ref{14}) with the deterministic evolution equations of
sequential dynamics\cite{coolen 1988}\cite{coolen 1992-1}, beyond our
expect, it is clear that both case is identical completely. Obviously, the
sequential dynamics with one-spin flip is a special case of synchronous
dynamics. The origin of this equality are the premises of continuous Markov
process and the cut of terms with higher order in the Fokker-Planck-type
approach for master equations. From the Lindeberg continuous condition of
Markov process\cite{Gardiner 1983}, the differentia between the
probabilities of contiguous states goes to zero faster, as the time step
goes to zero. To the extent that there only exist one or several updating
spin. As a result, the macroscopic parallel dynamics is the same as
sequential dynamics in the limit case.

Another interrelated topic is finite-size effects in networks. In another
word, the problem is how large is a small system. As a example, in numerical
simulation, it is necessary to consider system size up to $N\simeq 3\times
10^4$ for calculating certain properties of Hopfield model\cite{kohring 1990}%
. This puzzle how to account for the finite-size of the networks and extract
useful information about the asymptotic, $N\rightarrow \infty $, limit form
networks of only a few hundred to a few thousand neurons, was studied by
Forrest\cite{forrest 1988}, Kanter and Sompolinsky\cite{Kanter 1987}.
Furthermore, in the asynchronous case, the more detail analysis exploration
was performed by Castellanos, Coolen and Viana basing on the CS theory\cite
{castellanos 1998}, with the result that the effects is governed by a
time-dependent Ornstein-Uhlenbeck process in the condition of away from
saturation. In the following context, we will investigate this effects of
systems in synchronous evolutionary circumstance.

Taking advantage of the above route, it is easy to get the quadric term of
the right side of Eq. (\ref{10})

\begin{eqnarray}
&&\sum\limits_{{\bf s}^{\prime }}\left[ {\bf \Omega }\left( {\bf s}^{\prime
}\right) -{\bf \Omega }\left( {\bf s}\right) \right] ^2w\left( {\bf s}%
\rightarrow {\bf s}^{\prime }\right)  \nonumber \\
&=&\frac 4{N^2}\left\{ \left[ \sum\limits_i^N\frac{{\bf \xi }_is_i-{\bf \xi }%
_i\tanh \left( \beta h_i\left( {\bf \Omega }\right) \right) }2\right]
^2+\sum\limits_i^N\frac{{\bf \xi }_i^2\left( 1-\tanh ^2\left( \beta
h_i\left( {\bf \Omega }\right) \right) \right) }4\right\}  \nonumber \\
&=&\left[ {\bf \Omega -}\frac 1N\sum\limits_i^N{\bf \xi }_i\tanh \left(
\beta h_i\left( {\bf \Omega }\right) \right) \right] ^2+\frac 1{N^2}%
\sum\limits_i^N{\bf \xi }_i^2\left[ 1-\tanh ^2\left( \beta h_i\left( {\bf %
\Omega }\right) \right) \right]  \label{15}
\end{eqnarray}
It is straightforward to obtain the following Fokker-Planck-type equation
from substituting the Eq. (\ref{15}) into Eq. (\ref{10}): 
\begin{eqnarray}
\frac{dP_t\left( {\bf \Omega }\right) }{dt} &=&\sum\limits_k\frac \partial {%
\partial \Omega _k}\left\{ P_t\left( {\bf \Omega }\right) \left[ \Omega
_k\left( {\bf s}\right) -\frac 1N\sum\limits_i^N{\bf \xi }_i\tanh \left(
\beta {\bf \xi }_i{\bf \cdot A\Omega }\right) \right] \right\}  \nonumber \\
&&+\frac 12\sum\limits_{k,l}\frac{\partial ^2}{\partial \Omega _k\partial
\Omega _l}\left\{ P_t\left( {\bf \Omega }\right) \left[ -\left( {\bf \Omega -%
}\frac 1N\sum\limits_i^N{\bf \xi }_i\tanh \left( \beta {\bf \xi }_i{\bf %
\cdot A\Omega }\right) \right) ^2\right. \right.  \nonumber \\
&&\left. \left. -\frac 1{N^2}\sum\limits_i^N{\bf \xi }_i^2\left( 1-\tanh
^2\left( \beta {\bf \xi }_i{\bf \cdot A\Omega }\right) \right) \right]
\right\}  \label{16}
\end{eqnarray}
Now, the diffusion parameter is different to that in sequential dynamics\cite
{castellanos 1998}. It means both dynamics are more unlike in fluctuation or
internal noise distribution.

Following awake of the route of Castellanos {\it et. al.}\cite{castellanos
1998}. the new rescaled variable ${\bf q}$ and its probability distribution
function is defined as: 
\begin{equation}
{\bf q}\left( t\right) =\sqrt{N}\left( {\bf \Omega }\left( t\right) -{\bf %
\Omega }^{*}\left( t\right) \right) \qquad P_t\left( {\bf q}\right) =\int d%
{\bf \Omega }P_t\left( {\bf \Omega }\right) \delta \left( {\bf q}-\sqrt{N}%
\left( {\bf \Omega }-{\bf \Omega }^{*}\right) \right)  \label{17}
\end{equation}
where ${\bf \Omega }^{*}$ is the deterministic solution of Louville equation
(\ref{13}), and Eq. (\ref{17}) means that the order vector ${\bf \Omega }$
can be resolved into the sum of a deterministic terms ${\bf \Omega }^{*}$
and a fluctuating term with the latter terms vanishing in the limit of $%
N\rightarrow \infty $. Moreover, from the central limit theorem, the
fluctuating term can be scaled as $N^{-\frac 12}$\cite{shiino 1990}\cite
{scacciatelli 1992}.

In the limit of $N\rightarrow \infty $, with the help (\ref{16}), the
Fokker-Planck-type equation of rescaled variables is deduced:.

\begin{equation}
\frac{dP_t\left( {\bf q}\right) }{dt}=-\sum\limits_k\frac \partial {\partial
q_k}\left\{ P_t\left( {\bf q}\right) F_k\left( {\bf q},t\right) \right\} +%
\frac 12\sum\limits_{k,l}\frac{\partial ^2}{\partial q_k\partial q_l}\left\{
P_t\left( {\bf q}\right) D_{kl}\left( {\bf q},t\right) \right\}  \label{18}
\end{equation}
In there, the drift factor is given by 
\begin{eqnarray}
{\bf F}\left( {\bf q},t\right) &=&\beta \left\langle {\bf \xi }\left( {\bf %
\xi }\cdot {\bf Aq}\right) \left[ 1-\tanh ^2\left( \beta {\bf \xi }\cdot 
{\bf A\Omega }^{*}\right) \right] \right\rangle _{{\bf \xi }}-{\bf q} 
\nonumber \\
&&+\lim\limits_{N\rightarrow \infty }\sqrt{N}\left\{ \frac 1N\sum\limits_i%
{\bf \xi }_i\tanh \left( \beta {\bf \xi }_i\cdot {\bf A\Omega }^{*}\right) -%
{\bf \Omega }^{*}\right\}  \label{19}
\end{eqnarray}
and we denote $\left\langle g\left( {\bf \xi }\right) \right\rangle _{{\bf %
\xi }}=\lim_{N\rightarrow \infty }\frac 1N\sum_kg\left( {\bf \xi }_k\right) $%
which ${\bf \xi }_k=\left( \xi _k^1,\ldots ,\xi _k^p\right) $.The last term
of the right side of Eq. (\ref{19}) pictures a finite-size corrections to
the flow field. Similarly, the diffusion factor can be drawn 
\begin{eqnarray}
{\bf D}\left( {\bf q},t\right) &=&-\left\langle {\bf \xi }^2\left( 1-\tanh
^2\left( \beta {\bf \xi \cdot A\Omega }^{*}\right) \right) \right\rangle _{%
{\bf \xi }}  \nonumber \\
&&+\left\{ \left[ \beta \left\langle {\bf \xi }\left( {\bf \xi }\cdot {\bf Aq%
}\right) \left[ 1-\tanh ^2\left( \beta {\bf \xi }\cdot {\bf A\Omega }%
^{*}\right) \right] \right\rangle _{{\bf \xi }}-{\bf q}\right] \right. 
\nonumber \\
&&\left. +\lim\limits_{N\rightarrow \infty }\sqrt{N}\left[ \frac 1N%
\sum\limits_i^N{\bf \xi }_i\tanh \left( \beta {\bf \xi }_i{\bf \cdot A\Omega 
}^{*}\right) -{\bf \Omega }^{*}\right] \right\} ^2
\end{eqnarray}

Obviously, not as the system with sequential dynamics, the finite-size
effects for parallel dynamics is govern by a homogeneous Markov process. The
More intensive and detail work including verification of numerical
simulation is ongoing.

In short, the macroscopic dynamics of networks is all the same to both
updating ways, synchronous and asynchronous, in the limit of far away of
saturation. But there exist deviation in the fluctuation of pattern overlap
implied in Eq. (\ref{16}). Moreover, in the condition of away from
saturation, the finite-size effects in parallel dynamics is described by
homogeneous Markov process which is not as the time-dependent
Ornstein-Uhlenbeck process in sequential dynamics.

\end{document}